\documentclass[aps,amsmath,superscriptaddress]{revtex4-1}
\usepackage[pdftex]{graphicx}
\usepackage[figuresright]{rotating}
\usepackage{braket}
\usepackage{bm}
\usepackage{amsmath,amssymb}
\usepackage{color}
\usepackage{url}
\usepackage{hyperref}
\usepackage{algorithm}
\usepackage{algpseudocode}
\usepackage[titletoc]{appendix}
\usepackage{bbm}
\usepackage{bbold}
\usepackage{mathrsfs}

\begin{document}
\title{Propagating large open quantum systems towards their steady states: 
cluster implementation of the time-evolving block decimation scheme.}

\author{Valentin~Volokitin$^{1}$, Ihor~Vakulchyk$^{2,3}$, Evgeny~Kozinov$^{1}$, Alexey~Liniov$^{1}$, Iosif~Meyerov$^{1}$, Michail~Ivanchenko$^{1}$, Tatyana~Laptyeva$^{1}$, Sergey~Denisov$^{4}$}
\address{$^{1}$Lobachevsky State University of Nizhny Novgorod, Russia \\
$^{2}$Institute for Basic Science, Daejeon, Korea\\
$^{3}$Korea University of Science and Technology, Daejeon, Korea\\
$^{4}$Oslo Metropolitan University, N-0130 Oslo, Norway
}
%\ead{valentin.volokitin@itlab.unn.ru}
%%\author{V.~Volokitin}
%%\affiliation{Mathematical Software and Supercomputing Technologies Department, Lobachevsky State University of Nizhny Novgorod, Russia}
%%\author{I.~Vakulchyk}
%%\affiliation{Center for Theoretical Physics of Complex Systems,
%%Institute for Basic Science(IBS), Daejeon, Korea, 34126}
%%\affiliation{Basic Science Program, Korea University of Science and Technology(UST), Daejeon, Korea, 34113}
%%\author{E.~Kozinov}
%%\affiliation{Mathematical Software and Supercomputing Technologies Department, Lobachevsky State University of Nizhny Novgorod, Russia}
%%\author{A.~Liniov}
%%\affiliation{Software Engineering Department, Lobachevsky State University of Nizhny Novgorod, Russia}
%%\author{I.~Meyerov}
%%\affiliation{Mathematical Software and Supercomputing Technologies Department, Lobachevsky State University of Nizhny Novgorod, Russia}
%%\author{M.~Ivanchenko}
%%\affiliation{Department of Applied Mathematics, Lobachevsky State University of Nizhny Novgorod, Russia}
%%\author{T.~Laptyeva}
%%\affiliation{Department of Applied Mathematics, Lobachevsky State University of Nizhny Novgorod, Russia}
%%\author{S.~Denisov}
%%\affiliation{Department of Computer Science, Oslo Metropolitan University, N-0130 Oslo, Norway}

%\affiliation{Institut f\"ur Physik, Universit\"at Augsburg, Universit\"atsstra{\ss}e 1, D-86135 Augsburg, Germany}
%\affiliation{Department of Applied Mathematics, Lobachevsky State University of Nizhny Novgorod, Russia}

\begin{abstract}
Many-body quantum systems are subjected to the Curse of Dimensionality: The dimension of the Hilbert space $\mathcal{H}$,
where these systems live in,  grows exponentially with number of their components ('bodies'). 
%In order to specify a state of a quantum system, we need a description whose length grows exponentially with the system size.
However, with some systems it is possible to escape  the curse by using low-rank tensor approximations known as ``matrix-product state/operator (MPS/O)
representation'' in the quantum community and ``tensor-train decomposition'' among applied mathematicians.
%, which restricts the growth of the system's state description with $N$ to a linear law.
Motivated by recent advances in computational quantum physics, we consider chains of $N$  spins coupled by nearest-neighbor interactions.
The spins are subjected to an action coming from the environment. Spatially disordered interaction and environment-induced decoherence drive systems into non-trivial asymptotic  states. 
The dissipative evolution is modeled with a Markovian master equation in the Lindblad form.
By implementing the MPO technique and propagating system states with the time-evolving block decimation (TEBD) scheme (which allows keeping the length of the state descriptions fixed), 
it is in principle possible to reach the corresponding steady states. 
We propose and realize a cluster implementation of this idea. The implementation on four nodes allowed us 
to resolve steady states of the model systems with  $N = 128$ spins (total dimension of the Hilbert space $\mathrm{dim}\mathcal{H}  = 2^{128} \approx 10^{39}$).
\end{abstract}

\maketitle
    
\section{Introduction}\label{sec:1} 
Many-body systems are at the focus of the current research in theoretical and experimental quantum physics. 
In addition to their fundamental importance for  quantum thermodynamics and information \cite{nature}, these systems are perspective from 
the technological point of view; e.g., all manufactured (by now) quantum computers are based on arrays of interacting superconducting qubits \cite{tech2}).

All real-life  quantum systems are open, meaning that they interact -- to a different extent --
with their environments \cite{book}. This 'action from outside', termed ``decoherence`` or ''dissipation``, 
works together with the unitary evolution stemming from system's Hamiltonians and, on large time scales,
these joint efforts result in the creation of an asymptotic stationary (steady) state. The evolution of an open quantum system towards its steady states
is usually modeled with a Markovian master equation, which describes the dynamics of the system density operator $\varrho (t)$, 
$\dot{\varrho}(t) = \mathcal{L}\varrho (t)$ \cite{book}. Formally, similar to the Schroedinger equation used to describe unitary evolution of an isolated quantum system, this is a linear differential equation which can be solved numerically, e. g., by diagonalizing generator of evolution $\mathcal{L}$.

However, computational studies of many-body quantum systems are  limited by the so-called Course of Dimensionality: 
the total length $L$ of description (number of parameters required to specify a state) of an isolated  quantum system consisting of  
$N$ components (spins, qubits, ions, etc.), each one with $d$ degrees of freedom,  scales as $L(N) \sim d^{N}$. To specify an \textit{arbitrary} state of a system of $50$ qubits
one needs $2^{50} \approx 10^{15}$ complex-valued parameters. This exceeds the memory capacity of the supercomputer ``Titan''\cite{titan}.
In the case of open quantum systems, the complexity squares: to describe a density operator one needs $L(N) \sim d^{2N}$ real-valued parameters.

This is a famous problem in modern data science -- manipulations (or even simply storing) with data tensors becomes impossible when the data are sorted in high-dimensional spaces. The attempts to break the curse
led to the development of a variety of low-rank tensor approximation algorithms  \cite{tensor_decompose_review}. 
These algorithms are used now in signal processing,  computer vision,  data mining, and
neuroscience \cite{tensor1}. The most robust algorithms are based on  Singular Value Decomposition (SVD), 
and one particularly efficient for multilinear algebra manipulations is the so-called Tensor-Train (TT) decomposition \cite{oseled}. 
In physical literature, it is commonly referred to as Matrix Product State (MPS) [or Matrix Product Operator (MPO)] representation \cite{sh,schollw}. 
While these two names are used simultaneously (though in different fields), the underlying mathematical structure is the same \cite{hag}. 
The MPS/MPO/TT approach allows to reduce descriptions of \textit{some} many-body states to a linear scaling $L(N) \sim N$\cite{oseled}.

The MPS/MPO representation allows for effective propagation of quantum many-body systems in time by using the so-called Time-Evolving Block Decimation (TEBD) 
scheme \cite{vidal}. In short, this is a procedure to reduce the description of the state, obtained after every propagation step, to a given fixed length $L_{\mathrm{cut}}$. 
The accuracy of the propagation is controllable through $L_{\mathrm{cut}}$: If the information is thrown out after the restriction is substantial, the used TEBD propagation is bad and leads to a wrong description. Otherwise, it is good.
Some many-body systems 'behave' well during the TEBD propagation and so the amount of the neglected information is tolerable  (we are not going to discuss physical properties underlying 
such a 'good behavior' 
and refer the reader to an extensive literature on the subject; see. e.g., Ref.~\cite{schollw}). 
Important is that the MPO/TT-TEBD scheme can be used to propagate open systems \cite{ver} 
and thus get in touch with the corresponding steady states \cite{znid,NN}.
It is crucial therefore to estimate computational resources needed for the realization of this program. Here we report the results of our studies in these directions.

%\section{Numerical Approach and Lindblad}

\section{The algorithm}

\subsection{Tensor-Train Decomposition}

Here we mainly follow works \cite{oseled} and \cite{sh}; for more details, we refer the interested reader to them.

We start with  a  $N$-dimensional 
complex-valued tensor $A^{i_1,i_2\ldots i_N}$ with $i_k = 1, 2, \ldots M$. 
By \textit{gluing} together indices $i_2, i_3 \ldots i_N$ we obtain a $M \times M^{N-1}$ matrix to 
which we apply then SVD (henceforth we use notation without Hermitian conjugation for the last matrix in the decomposition)
\begin{equation}
A\left[i_1;i_2\ldots i_N\right] = \sum_{\alpha_1, \alpha_2} U_1(i_1;\alpha_1)\Lambda_1(\alpha_1;\alpha_2)V_1(\alpha_2;i_2 \ldots i_N),
\end{equation}
where $U$ and $V$ are unitary matrices and $\Lambda \geq 0$ is a diagonal matrix with entries being singular 
values $\lambda_j$. We 
assume that singular values are sorted in descending order, $\lambda_1 \geq \lambda_2 \geq \ldots \lambda_r$. 
Using diagonal structure of $\Lambda$,$\Lambda_{\alpha_1,\alpha_2} = \lambda_{\alpha_1} \delta_{\alpha_1,\alpha_2}$, where $\lambda_j$ 
are singular values, and reshaping $U_1$ as a $1 \times d$ tensor indexed by $i_1$, we get
\begin{equation}
    A\left[i_1,i_2\ldots i_N\right] = \sum_{\alpha_1}\Gamma^{[1]i_1}_{\alpha_0,\alpha_1}\lambda^{[1]}_{\alpha_1}V_1(\alpha_1;i_2 \ldots i_N).
\end{equation}

Repeating the same ``reshape-SVD-reshape'' procedure for $V_1$ and continue further iteratively, we arrive at the TT representation,
\begin{equation}\label{general_MPS}
  A\left[i_1,i_2\ldots i_N\right] = \sum_{\alpha_1 \ldots \alpha_{N-1}}\Gamma^{[1]i_1}_{\alpha_0,\alpha_1}\lambda^{[1]}_{\alpha_1}\Gamma^{[2]i_2}_{\alpha_1,\alpha_2}\lambda^{[2]}_{\alpha_2}\ldots \Gamma^{[N-2]i_{N-1}}_{\alpha_{N-2},\alpha_{N-1}}\lambda^{[N-1]}_{\alpha_N-1}\Gamma^{[N]i_N}_{\alpha_{N-1},\alpha_N}.
\end{equation}

One may interpret this structure as a ``train''  (see Fig.~1a) of $\Gamma$'s that encode local structure in each dimension, 
and $\lambda$'s that quantify correlations between them. Each $\Gamma^{[k]}$ is an 
array of $M$ matrices $r_{k-1} \times r_k$ with restrictions $r_j\leq M \max{(r_{j-1},r_{j+1})}$ with 
boundary conditions $r_0=r_N=1$. Thus, the  dimensions of the matrices are $1 \times M, \; M \times M^2, \; M^2 \times M^3 \ldots M^2 \times M, \; M \times 1$, which 
corresponds to the full representation with $M^N$ complex parameters. 
%However, if during one of the stages SVD produces one or 
%more zero singular values $\lambda^{[p]_j}=0, \; j\geq l$ one cuts summation over $\alpha_p$ so that $r_p=l$ which provides effective 
%compression. However, exact zero singular values is an extremely rare occurrence in numerical applications. Luckily, TT provides a solid 
%basis for controlled error approximations. 
When SVD is performed, one can keep only certain singular values based on the approximation criterion. 
One possibility is to discard all values smaller than a fixed number. An alternative approach is to introduce a so-called \textbf{bond dimension} $R$,
a cut-off value such that on each bound $i$ only singular values $\lambda^{[i]}_j$, $j \leq R$, are kept and the rest are truncated. We use the latter option. 
Each local approximation procedure on the set of singular value $\{\lambda^{[i]}\}$ introduces a truncation error 
\begin{equation}\label{truncation_error_local}
     E_i(R) = \sum_{j>R} \left( \lambda^{[i]}_j \right)^2.
\end{equation}

One of the main advantages of the TT representation is the simplicity of local convolutions with other tensor structures. 
Consider an operation $T_j$ acting in $j$-th dimension only,
\begin{equation}
    A^\prime\left[i_1 \ldots i_j \ldots i_N\right] = \sum_{i^\prime_j,i_j}T_j[i_j,i^\prime_j]A\left[i_1 \ldots i^\prime_j \ldots i_N\right].
\end{equation}
After  substitution in Eq.~(\ref{general_MPS}), one could see that this convolution only affects the corresponding $\Gamma$ tensor,
\begin{equation}\label{one_index_op_map}
    \Gamma^{\prime[j]i_j} = \sum_{i_j^\prime}T_{i_j,i^\prime_j}\Gamma^{[j]i_j^\prime}.
\end{equation}
%

%%%%%%%%%%%%%% move to "Implementation" in case of need %%%%%%%%%%%%%%%%%%%%%%
\begin{figure}[h]
\includegraphics[width=20pc]{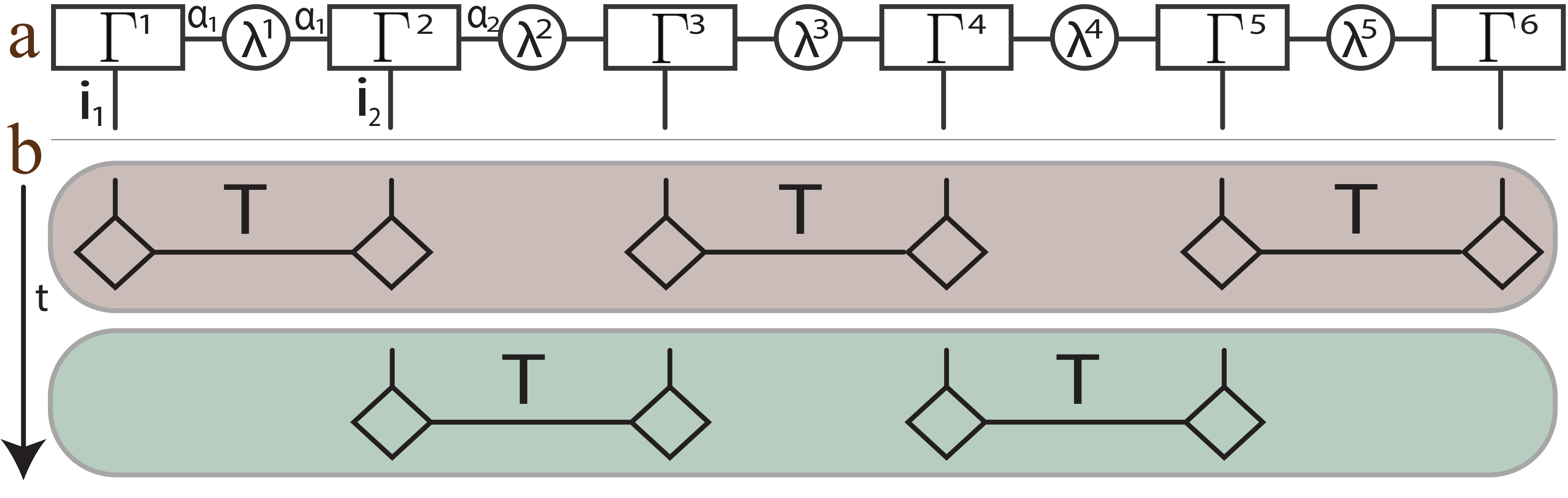}\hspace{2pc}%
\begin{minipage}[b]{18pc}\caption{\label{fig:01} (a) Tensor train (matrix product state) decomposition  and (b) Suzuki-Trotter propagation (see text for more details).}
\end{minipage}
\end{figure}

%\begin{figure}[h]
%\includegraphics[width=0.48\textwidth]{fig0}
%\caption{\label{fig:01} (a) Tensor train (matrix product state) decomposition  and (b) Suzuki-Trotter propagation (see text for more details).}
%\end{figure}
%%%%%%%%%%%%%%%%%%%%%%%%%%%%%%%%%%%%%%%%%%%%%%%%%%%%%%%%%%%%%%%%%%%%%%%%%%%%%%

Operations involving multiple dimensions (indices), especially distant,
destroy TT structure, so additional procedures are required to restore it. 
Generally, it would imply effectively the same procedure as initial decomposition. 
However, in this paper, we only use convolutions involving two neighboring indices. Thus, only one local reorthogonalization is required.

Consider an operation $T_{j,j+1}$:
\begin{equation}
A^\prime\left[i_1 \ldots i_j i_{j+1} \ldots i_N\right] = \sum_{i^\prime_j,i_j}T_{j,j+1}[i_j,i^\prime_j;i_{j+1},i^\prime_{j+1}] A\left[i_1 \ldots i^\prime_j i^\prime_{j+1}\ldots i_N\right].
\end{equation}
It affects a pair of the corresponding tensors,
\begin{equation}\label{two_index_op_map}
\Gamma^{[j]i_j}_{\alpha_{j-1}\alpha_j}, \; \lambda^{[l]}_{\alpha_j}, \; \Gamma^{[j+1]i_{j+1}}_{\alpha_j \alpha_{j+1}} \rightarrow 
 \sum_{i_j^\prime i_{j+1}^\prime, \alpha_j} T_{j,j+1}[i_j,i^\prime_j; i_{j+1},i^\prime_{j+1}] \Gamma_{\alpha_{j-1}\alpha_j}^{[j]i_j^\prime} \lambda^{[l]}_{i_j} \Gamma^{[j+1]i_{j+1}^\prime}_{\alpha_j \alpha_{j+1}}.
\end{equation}

To perform reorthogonalization, we 
introduce matrix  $K\left[i_{j},i_{j+1};\alpha_{j-1},\alpha_{j+1}\right]$. Next we reshape indexes, perform SVD, and reshape indexes back \cite{vidal_tebd}:
\begin{equation}\label{reorthog}
    K[\alpha_{j-1},i_j; \alpha_{j+1}i_{j+1}] = \Gamma^{\prime[j]i_j}_{\alpha_{j-1}\alpha_j} \lambda^{\prime[j]}_{\alpha_j}\Gamma^{\prime[j+1]i_{j+1}}_{\alpha_{j}\alpha_{j+1}}.
\end{equation}

Such operation takes only $\mathcal{O}(R^3)$ steps and this scaling does not depend on  $N$. 
Moreover, it modifies only a pair of relevant tensors so multiple pairwise operations that act in independent subspaces may be parallelized. 
The TT representation allows also for  fast realizations of  other algebraic operations: 
calculation of partial or full traces, norm, scalar products, additions, etc \cite{oseled}. 

\subsection{Tensor-Train Propagation}

%%%%%%%%%%%%%% move to "Implementation" in case of need %%%%%%%%%%%%%%%%%%%%%%
\begin{figure*}
\begin{minipage}{\linewidth}
%\begin{algorithm*}[htb]
\begin{algorithm}[H]
\caption{: TEBD method implementation}
\begin{algorithmic}[1]
    \State \textbf{upload}: system $\&$ method parameters (N, $T_{j}[i_{j},i'_{j}]$, $T_{j,j+1}[i_{j},i'_{j},i_{j+1},i'_{j+1}]$, $dt$, $T_{\rm{max}}$, $R$), initial state  ($\Gamma^{[j]i_j}_{\alpha_{j-1}\alpha_j}$, $\lambda^{[j]}_{\alpha_j}$)
    \For {$t=0$ \textbf{to} $T_{\rm{max}}$}
    \State propagate all $\Gamma^{[j]i_j}_{\alpha_{j-1}\alpha_j}$ on $[t; t+dt/2]$
    \State propagate $\Gamma^{[j]i_j}_{\alpha_{j-1}\alpha_j}$, $\lambda^{[j]}_{\alpha_j}$, $\Gamma^{[j+1]i_{j+1}}_{\alpha_{j}\alpha_{j+1}}$ with odd $j$ on $[t; t+dt/2]$  \Comment{4-6 do with hard cutoff of the local bond dimension}
    \State propagate $\Gamma^{[j]i_j}_{\alpha_{j-1}\alpha_j}$, $\lambda^{[j]}_{\alpha_j}$, $\Gamma^{[j+1]i_{j+1}}_{\alpha_{j}\alpha_{j+1}}$ with even $j$ on $[t; t+dt]$    
    \State propagate $\Gamma^{[j]i_j}_{\alpha_{j-1}\alpha_j}$, $\lambda^{[j]}_{\alpha_j}$, $\Gamma^{[j+1]i_{j+1}}_{\alpha_{j}\alpha_{j+1}}$ with odd $j$ on $[t+dt/2; t+dt]$ 
    \State propagate all $\Gamma^{[j]i_j}_{\alpha_{j-1}\alpha_j}$ on $[t+dt/2; t+dt]$
    %\State calculate $t_{\rm{next}}$
    \EndFor
    \State \textbf{save} results
    \State \textbf{release} memory
\end{algorithmic}
\end{algorithm}
\end{minipage}
\end{figure*}
%%%%%%%%%%%%%%%%%%%%%%%%%%%%%%%%%%%%%%%%%%%%%%%%%%%%%%%%%%%%%%%%%%%%%%%%%%%%%%

The TT representation provides a basis for an approximate tensor propagation algorithms. 
Here we use Time-Evolving Block Decimation (TEBD) scheme \cite{vidal_tebd, ver}, which was 
specifically designed for quantum systems but applicable also in the general case. Consider a tensor flow governed by an evolution generator 
consisting only of operations acting on one or two adjacent dimensions,
\begin{equation}
    \frac{d}{dt}A[i_1\ldots i_N]=\sum_j \sum_{i_j}T^{[1]j}_{i_j}A[i_1\ldots i_j \ldots i_N] + \sum_{j_1,j_2}\sum_{i_{j_1}, i_{j_2}}T^{[2]j}_{i_{j_1}, i_{j_2}}A[i_1\ldots i_{j_1} i_{j_2} \ldots i_N].
\end{equation}

We use standard time discretization to iteratively integrate this equation (starting from some initial tensor). 
In terms of operations the solution reads
\begin{equation}
    A(t+dt) = L(dt)A(t) = \\
        = \exp\left[\left(\sum_j \hat{T}^{[1]j} + \hat{T}^{[2]j} \right)dt\right]A(t).
\end{equation}

As $T$ operators generally do not commute, we have to approximate the matrix exponents. 
A way to minimize the error is to separate the operators into groups as large as possible such that all the operators belonging to one group commute with each other. 
All one-dimension operators commute by default, and two-dimension acting on odd/even pairs commute within their oddity groups; see Fig.~1b. 
We utilize this fact and use modified second order Suzuki-Trotter decomposition \cite{schollw}:
\begin{gather}\label{suzuki_trotter}
    L(dt) \approx L_1(dt/2)L_2^{\text{odd}}(dt/2)L_2^{\text{even}}(dt)L_2^{\text{odd}}(dt/2)L_1(dt/2), \nonumber \\
    L_1(dt) = \prod_{j=1}^N L^i_1(dt), \: L_2^{\text{even/odd}}(dt) = \prod_{j \in \{\text{even/odd}\}} L^i_2(dt), \\
    L^i_1(dt) = e^{\hat{T}^{[1]j} dt}, \; L^i_2(dt) = e^{\hat{T}^{[2]j} dt}.
\end{gather}

Note that the standard approach is to adsorb one-index operators $L_1$ into $L_2$, 
but we find it numerically beneficial to separate them. 
Each factor of $L_1$ and $L_2$ can be calculated by using Eqs.~ (\ref{one_index_op_map}) and (\ref{two_index_op_map}) respectively. 
Furthermore, as they commute by construction, corresponding computation can be parallelized. Each two-index operator may include a 
cut-off if after the reorthogonalizationm Eq.~(\ref{reorthog}), the number of singular value exceeds bound dimension $R$. Corresponding accumulated truncation 
error is then calculated as a sum of local errors (\ref{truncation_error_local}) over all the operation during evolution up to time $t$,
\begin{equation}\label{truncation_error}
    E(t, R) = \sum_{j=1}^{t/dt}\sum_{\{L^i_2(dt_j)\}} E^{(j)}_i(R).
\end{equation}
Computations are dominated by SVD, so resulting complexity is $\mathcal{O}(N R^3)$, where $R$ is the  bond dimension. 
With $\mathcal{O}(N)$ cores available, it becomes $\mathcal{O}(R^3)$ 
and thus the computational task is perfectly scalable.

\subsection{Lindblad Equation}
We apply both TT and TEBD methods to evolve numerically many-body open quantum models. 
The state of such systems is described by a density matrix $\varrho(t)$ of the size  $M^N \times M^N$, where $N$ is 
the number of particles/spins and $M$ is number of the local states, which we put to $M=2$ for a $1/2$-spins that we consider 
in the paper. Evolution of a quantum system in contact with the environment is governed by Lindblad equation \cite{book}
\begin{equation}\label{lindblad}
         \dot{\varrho}(t) = \mathcal{L}\varrho(t) = \mathcal{L}_H\varrho(t) + \mathcal{L}_{\mathrm{dis}}\varrho(t) = -i\left[H,\varrho(t)\right] + \sum_{s=1}^M \gamma_s \left[D_{s}\varrho(t)D^\dagger_{s}-\frac{1}{2}\{D^\dagger_{s}D_{s},\varrho(t)\}\right],
\end{equation}
where $\mathcal{L}$ is the Lindblad superoperator consisting of conservative $\mathcal{L}_H$ and dissipative $\mathcal{L}_{\mathrm{dis}}$ parts,
$H$ is Hamiltonian, $D_s$ are dissipation operators, and $\gamma_s$ are corresponding dissipation rates. 
There is a stationary state solution for any Lindblad superoperator $\mathcal{L} \varrho(\infty) = 0$ which is unique (aside of special cases of symmetries which we do not address here).

Many-body density operator $\varrho$ can be represented as an 
$2N$-dimensional tensor $\varrho \left[i_1,i_2\ldots i_N; i_1^\prime,i_2^\prime\ldots i_N^\prime\right]$ 
where every pair of indexes $i_j, i_j^\prime$ (each one runs from $1$ to $2$) correspond to the $j$-th qubit/spin. 
The models we consider include only one- and two-particle interactions, 
which thus involve up to four indexes of $\varrho$. 
To overcome this, we use vectorization procedure \cite{ver} and glue together 
indexes of each particle forming $\varrho\left[i_1 i_1^\prime; i_2 i_2^\prime \ldots i_N i_N^\prime \right]$ -- $N$-dimension 
tensor with each dimension going from $1$ to $4$. This allows applying TEBD scheme (\ref{suzuki_trotter}) as long as $H$ and $A_s$ in (\ref{lindblad}) do not couple more than two particles. 
However, as we restrict the accessible space to the bond dimension  $R$, 
evolution can only start from initial MPO conditions belonging to this space. In the models considered further, we use an extreme case of product initial states, $r_j=1$,  $\forall j$;
as we aim at the stationary states,  we have complete freedom when choosing initial conditions 
(though other choices can be more beneficial from the relaxation-speed point of view).

\section{Model system}\label{sec:2} 

As test-bed models, we use spin chains from Refs.~\cite{znid,NN}. Here we only briefly described them. 

Both chains consist of $N$ spins. Hamiltonian part of evolution is governed by the Heisenberg XXZ model with local disordered potential
\begin{equation}
    H=\sum_i \sigma^x_i \sigma^x_{i+1} + \sigma^y_i \sigma^y_{i+1} + J \sigma^z_i \sigma^z_{i+1} + h_i \sigma^z_i, 
\end{equation}
where $\sigma_i$ are Pauli matrices, $J$ is interaction strength parameter and $h_i$ are uncorrelated random value uniformly distributed in $[-h,h]$. Open boundary conditions are used.

In Ref.~\cite{znid}, a disordered  spin chain  with next-neighbor coupling and 
two “thermal reservoirs” -- each one represented by  pair of Lindblad operators 
causing excitation (relaxation) and acting on the two end spins, $i = 1,N$, -- were considered,
\begin{equation}
    D_\pm^{i=1} = \sqrt{1 \pm \mu}\sigma^\pm_1, \; D_\pm^{i=N} = \sqrt{1 \mp \mu}\sigma^\pm_N,
\end{equation}
where $\mu$ is bias responsible for the formation of non-equilibrium steady state with directed non-zero current. 
In the limit $\mu \rightarrow 0$, the stationary state is an infinite temperature '(maximally mixed') state, that is the normalized identity. 
By assuming $\mu \ll 1$, one could address the linear-response regime. The transport of the spin charge through a chain in the stationary regime was considered 
and the current scaling with $N$ was estimated.

The authors managed to achieve a model size of $N=400$, which is an unprecedented size for many-body open quantum models. The complexity of the computational experiments is increased by
the fact that the systems needed a considerable propagation time in order to reach the stationary state (this is because the dissipation was acting at the chain ends only). Finally, to obtain scaling dependencies, 
averaging over disorder realizations was performed. 
At the same time, this work provides little detail about the resources used for numerical simulations. 
We considered the obtained results as a challenge and decided to reproduce them -- at least for $N \simeq 100$.

As an additional test-bed, we use a model from another recent work \cite{NN}. 
In this paper, disordered spins chains in which all spins are subjected to the action of dissipation were considered. 
Dissipative terms couple each pair of spins,
\begin{equation}
    D_l = (\sigma_{l}^+ + \sigma_{l+1}^+)(\sigma_{l}^--\sigma_{l+1}^-),
\end{equation}
which try to make neighboring spins oscillating out of phase (`anti-synchronization').
The maximal size of the models used in numerical simulations, reported in this work, was $N=32$. 
Among other characteristics, scaling of the so-called operational entanglement entropy ~\cite{Prosen2007} was considered. 
We use this quantifier in our numerical experiments, in which we tried to reach $N=128$.

\section{Implementation}\label{sec:3} 
The method described in Section II is implemented as shown in Algorithm 1.
The algorithm is implemented using the C++ programming language. We found, that the matrix operations (mainly SVD) are the most time-consuming parts of the algorithm. 
In this regard, we employ the Armadillo software library integrated with highly optimized mathematical routines from the Intel Math Kernel Library to improve performance. Finally, Armadillo/MKL routines take about 50-80\% of computation time during the propagation step depending on the current system state.

The algorithm assumes performing a set of integration operations for individual components of the system at every time step. These operations are not independent but can be ordered according to their dependencies for the organization of parallel computations. In particular, all one- and two-particle interactions can be performed completely in parallel.

\begin{figure}[h]
\includegraphics[width=20pc]{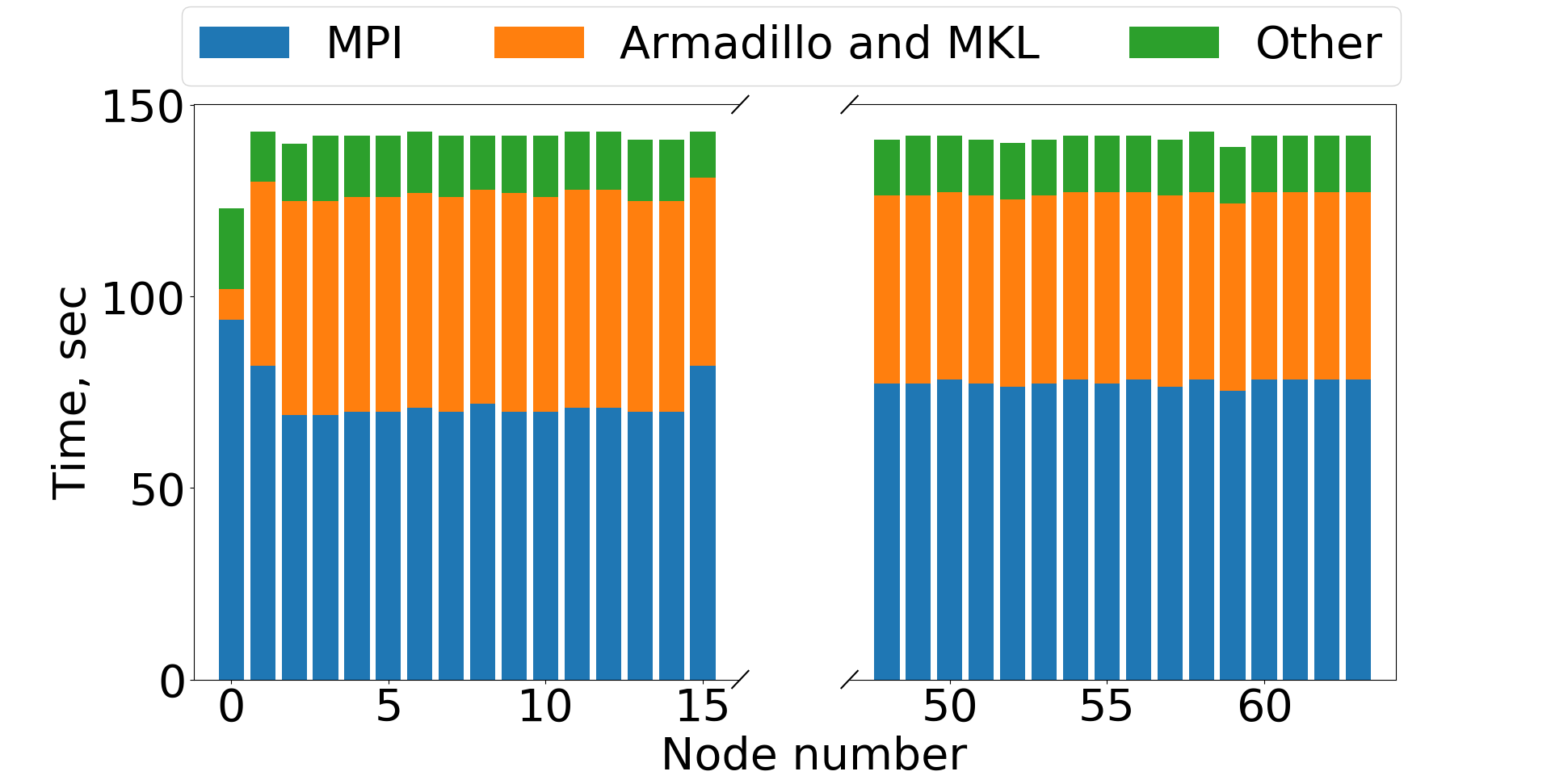}\hspace{2pc}%
\begin{minipage}[b]{14pc}\caption{\label{fig:timesnew}Distribution of computational and communication functions run time. 64 MPI-processes were executed on four nodes of the cluster.}
\end{minipage}
\end{figure}

%\begin{figure}
%    \includegraphics[width=0.47\textwidth]{Times_new}
%    \caption{\label{fig:timesnew}Distribution of computational and communication functions run time. 64 MPI-processes were executed on four nodes of the cluster.}
%\end{figure}

The cluster parallelization is done by using the MPI technology. 
We apply the classic master-worker scheme for parallelization of the algorithm. 
For that, the single managing MPI-process (master) forms separate tasks for one-- and two--particle interactions, 
monitors their dependencies from each other and readiness, distributes tasks to all other processes (workers) and accumulates the results.

All computational experiments have been done on the Lobachevsky cluster with a $2 \times8$-core Intel Xeon CPU E5-2660, 2.20GHz, 
64 GB RAM, Infiniband QDR interconnect. 
The code was compiled with the Intel C++ Compiler, Intel Math Kernel Library and Intel MPI from the Intel Parallel Studio XE suite of development tools and the Armadillo library.

We integrate model from Ref.~\cite{znid} with following parameters: $N = 128$, $R = 50$, $T_{\rm{max}} = 50$, $dt = 0.1$. 
Parallel code was run on four computational nodes of the cluster (1 MPI-process per CPU core, 64 MPI-processes overall). Total computation time was 143~s. The resulted diagram for the distribution of computational and communication functions run time is presented in Fig.~\ref{fig:timesnew}. It is shown that the calculations are fairly well balanced, which is an undoubted advantage of the parallelization scheme. However, MPI communications take a significant part of the computation time, while further increasing the number of cluster nodes used will not significantly speed up the calculations, which is a limitation of the scheme. Computational efficiency (ratio of computation time to total execution time) was 47\%.

%To check correctness of the performed experiment, we integrated the system with a different number of allowed SVD 
%numbers/vectors $R$ = 60, 90, 120, 240, 360, 480. The sum of the modules of all the dropped SVD numbers was calculated. T
%he results are presented in Fig.~\ref{fig:truncatederror}.

\begin{figure}[h]
\includegraphics[width=18pc]{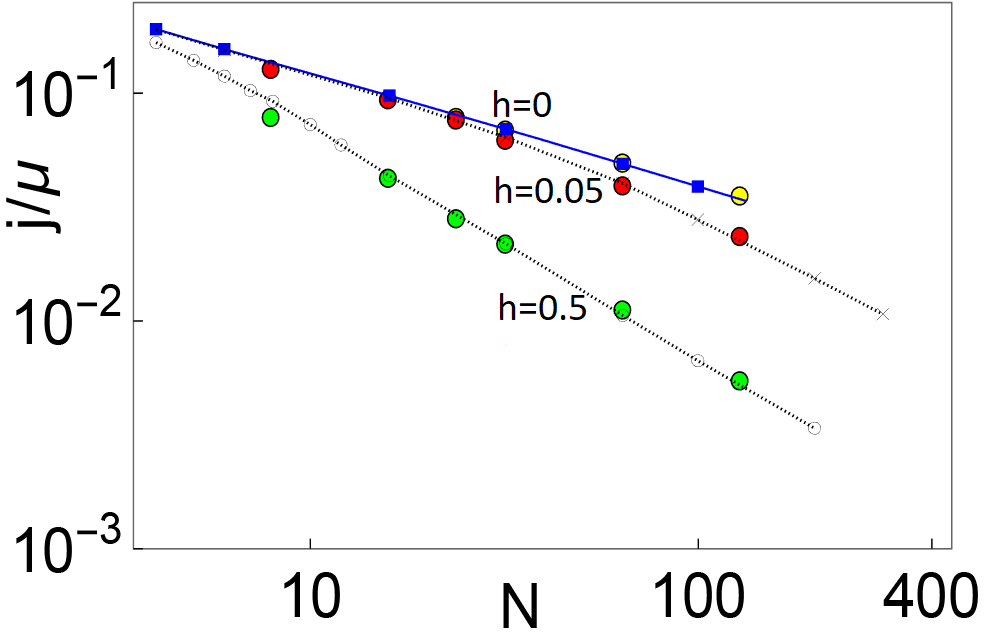}\hspace{2pc}%
\begin{minipage}[b]{18pc}\caption{\label{fig:truncatederror} Scaling of the spin current $j$ through a disordered spin chain with $N$ spins for different values of disorder strength $h$. 
Our results (big colored circles) are plotted on top of the results (lines and other symbols) reported in  Ref.~\cite{znid}). The maximal size of the model system used in our simulations is $N = 128$.
For every set of parameters, we performed averaging over $20$ disorder realizations. The propagation time step  $dt=0.1$ and bond dimension $R=50$.}
\end{minipage}
\end{figure}

%\begin{figure}[t]
%\includegraphics[width=0.45\textwidth]{1_2}
%\caption{\label{fig:truncatederror} Scaling of the spin current $j$ through a disordered spin chain with $N$ spins for different values of disorder strength $h$. 
%Our results (big colored circles) are plotted on top of the results reported in  Ref.~\cite{znid}). The maximal size of the model system used in our simulations is $N = 128$.
%For every set of parameters, we performed averaging over $20$ disorder realizations. The propagation time step  $dt=0.1$ and bond dimension $R=50$.}
%\end{figure}

%\begin{itemize}
%\item[]  First serial code: all the CPU time is spent on SVD and 
%(much less) building theta-matrices for SVD. Discussion of complexity, $O(d^3*N)$, or similar. Discussion of (failed) attempts to optimize SVD. MKL is used as the best tool available.

%\item[] First parallel code: each core computes propagation of a pair of sites one in a time, then the state data are redistributed, etc.

%\item[] Analyses of the errors, discussion of the two approximation parameters: time step $dt$ and cut-off bond dimension $\chi$. 

%\item[] Potential improvement: code is still not fully parallel/scalable as after each time step there is a serial part -- 
%calculation of the observables/physical values on the master core. Should be easy to solve by, for example, saving the state 
%and creating a single-core task for calculating these values without stopping calculations

%\item[] Resource scaling
%\end{itemize}

\section{Results}\label{sec:4} 

We find that it is possible to reproduce - with high accuracy -- the results reported in Ref.~\cite{znid}  by using bond-dimension $R=50$.
On Figure. 3 we present a comparison of the results of the sampling we perform with our code (big circles; yellow, red and green) with the results by \v{Z}nidari\u{c} and his co-authors.
We use propagation step $dt = 0.1$ and propagate every system up to $t = 10^4$, irrespectively of its size. It is not the most optimal way to sample (for example, 
it would be more effective  to determine arrival of the system at the asymptotic state  by monitoring the value of the spin current); however, at this stage, we tried
to make the sampling procedure as simple as possible.
For every value of $N$ and disorder strength $h$, we additionally performed averaging for $20$ disorder realizations. 
Each realization took from $2$ minutes to $2$ hours depending on the size of the system and involved up to two nodes (for large system sizes, $N > 64$).

For the model from Ref.~\cite{NN} we study the dynamics of the operator entanglement entropy (for a fixed disorder realization) 
for different bond dimensions. We found that $R_c = 360$  constitutes a threshold value after which the asymptotic entropy does not change upon further increase of the bond dimension. The calculation time for this value of bond dimension was four weeks of continuous propagation on four computing nodes.

We analyze also the evolution of accumulated error \ref{truncation_error} in this case. It is noteworthy that saturation of the operator entropy, which signals the arrival to the asymtotic state, is not accompanied by the saturation of the error. The latter continues to grow in a power-law manner, see the dashed line in Fig.~4b. 
This means that MPO states -- even with $R = 480$ -- are different from the genuine steady state of the model [which is the zero-value eigenelement of 
the corresponding Lindbladian (16 - 18)].

\section{Conclusions}\label{sec:5}

We presented a parallel implementation of the MPO-TEBD algorithm to propagate many-body open quantum systems. 
Parallelization is performed using the MPI technology and employs the master-worker scheme for computational tasks distribution. 
High-performance implementations of linear algebra from the Intel MKL were used to better utilize computational resources of modern hardware. 

A series of numerical experiments was performed to determine the accuracy and limits of applicability of the developed code. 
In particular, the effect of the number of SVD numbers kept after each propagation step (bond-dimension $R$) on the accuracy of the method was investigated. 
We found that threshold value $R_{c}$ after which saturation of the relevant characteristics is observed and fuhrer increase of bond dimension does not change their values.

The performance tests on the Lobachevsky cluster demonstrated that $64$ MPI processes running on four computational nodes is the optimal configuration for the model systems with $N=128$ spins.
As a next step, we plan to explore the possibility of further improvements of the parallelization by reducing the communications and increasing the efficiency of using computational resources.
After that, we hope to reach the limit $N \simeq 400$ with the test-bed models.

\begin{figure}
    \includegraphics[width=0.65\textwidth]{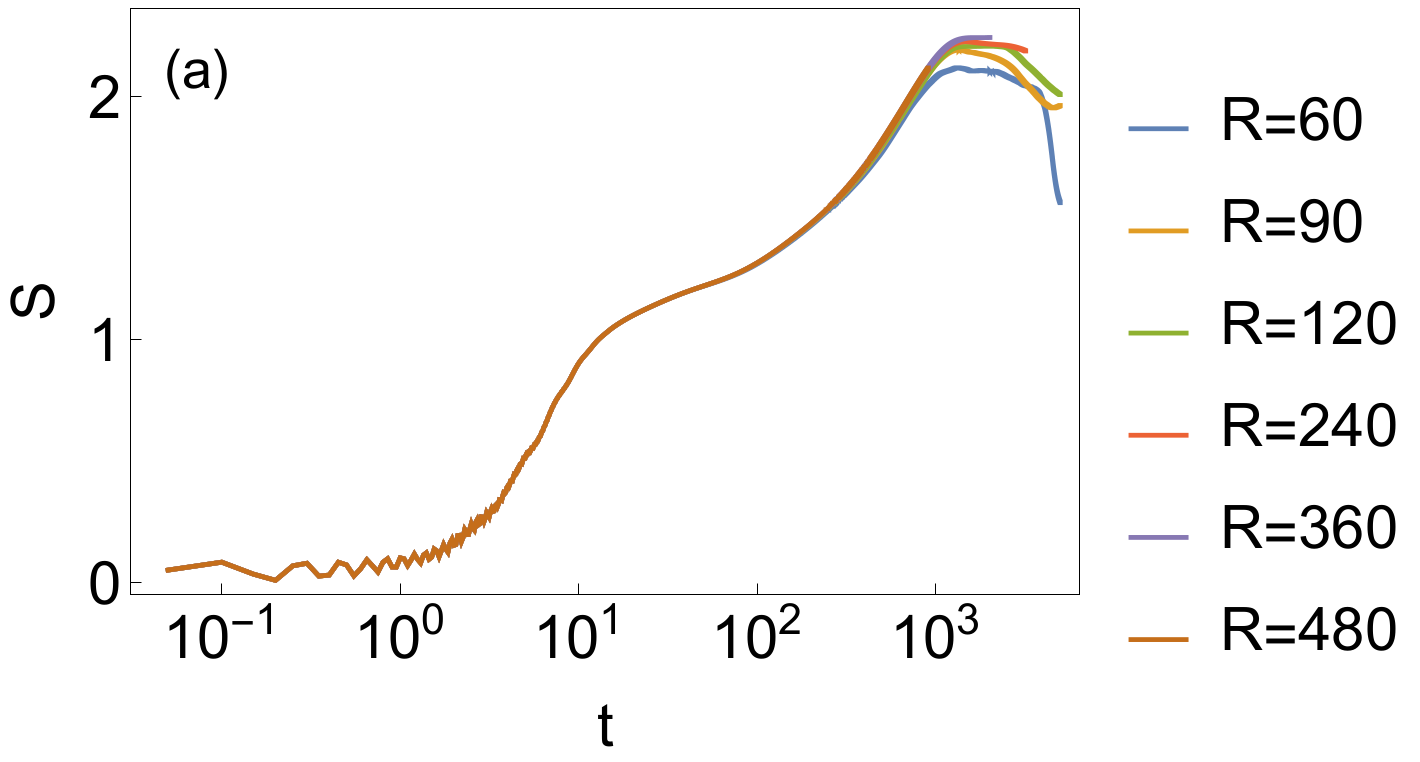}
    \includegraphics[width=0.65\textwidth]{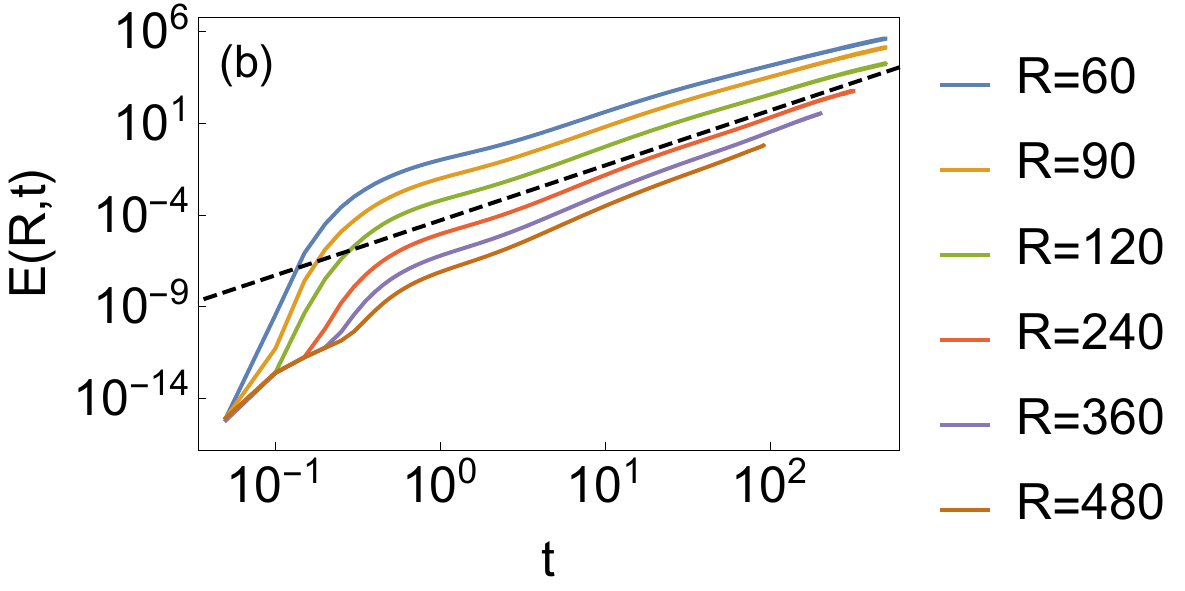}
    \caption{\label{fig:truncatederror} (a) Evolution of the operator entanglement entropy $S$ for a single disorder realization for the model from Ref.~\cite{NN}, 
    for different values of bond dimension $R$. The propagation time step  $dt=0.1$ and the system size is $N=128$. Note that for $R=480$ we did not reach the asymptotic 'plateau' because
    it was not possible to numerically propagate system further (we hit the two-week limit).
    (b) Increase of the accumulated truncation  error (\ref{truncation_error}) in time. 
    }
\end{figure}

    \section{Acknowledgments}\label{acknowledgment}
    The authors acknowledge support of the Russian Foundation for Basic Research and the Government of the Nizhni Novgorod region of the Russian Federation, grant \#~18-41-520004. IV acknowledges support by the  Institute for Basic Science, Project Code (IBS-R024-D1), and by the Korea University of Science and Technology Overseas Training program.

\end{document}